\documentclass[final]{cvpr}

\usepackage{times}
\usepackage{epsfig}
\usepackage{graphicx}
\usepackage{amsmath}
\usepackage{amssymb}
\usepackage{float}
\usepackage{array}
\usepackage{tabularx}
\usepackage{subfig,graphicx}

\usepackage[pagebackref=true,breaklinks=true,colorlinks,bookmarks=false]{hyperref}

\begin{document}

\title{Spatio-Temporal Graph Convolutional Network combined Large Language Model: A Deep Learning Framework for Bike Demand Forecasting}

\author{Peisen Li\\
Weiyang College\\Tsinghua University\\
{\tt\small lps20@mails.tsinghua.edu.cn}
\and
Yizhe Pang\\
Weiyang College\\ Tsinghua University\\
{\tt\small pangyz20@mails.tsinghua.edu.cn}
\and
Junyu Ren\\
Department of Automation\\ Tsinghua University\\
{\tt\small renjy22@mails.tsinghua.edu.cn}
}

\maketitle

\begin{abstract}
This study presents a new deep learning framework, combining Spatio-Temporal Graph Convolutional Network (STGCN) with a Large Language Model (LLM), for bike demand forecasting. Addressing challenges in transforming discrete datasets and integrating unstructured language data, the framework leverages LLMs to extract insights from Points of Interest (POI) text data. The proposed STGCN-L model demonstrates competitive performance compared to existing models, showcasing its potential in predicting bike demand. Experiments using Philadelphia datasets highlight the effectiveness of the hybrid model, emphasizing the need for further exploration and enhancements, such as incorporating additional features like weather data for improved accuracy.
\end{abstract}

\section{Introduction}
Bicycle sharing is an important part of the transportation and life of citizens in the city. Since the flow of people in the whole city varies greatly in different locations, it has become an important issue to allocate shared bicycles in the whole city according to the traffic flow. Specifically, we need to predict the distribution relationship of shared bicycles in the entire spatio-temporal domain according to the existing traffic flow of the whole city, so as to help relevant practitioners and staff to carry out the pre-allocation of shared bicycles\cite{liangBikeSharingDemand2022a,liangJointDemandPrediction2022}.

\begin{figure}[t]
\begin{center}
\fbox{
   \includegraphics[width=0.8\linewidth]{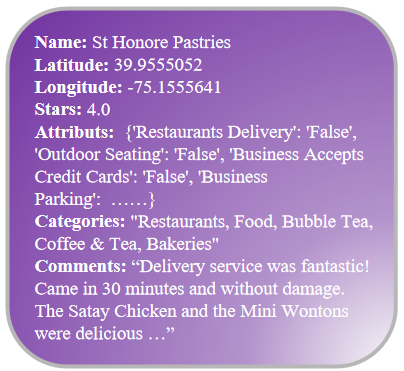}
}
\end{center}
   \caption{Yelp Business POI data example}
\label{fig:yelp}
\end{figure}

The main problem of the shared bicycle prediction task is how to convert the discrete dataset into continuous and highly spatiotemporally dependent data, so as to construct a deep learning framework with practical application value.

According to the available data, the traffic flow of shared bicycles in cities is often discrete. Traditional structured mobility models are unable to handle this unstructured language data, resulting in poor performance of shared bicycle prediction tasks.

Contemporary challenges encountered by mobility models encompass two principal impediments in harnessing language data effectively. Firstly, they inadequately exploit the extensive reservoirs of descriptive text pertaining to Points of Interest (POI). Secondly, the integration of unstructured text data into structured mobility models presents a formidable challenge. Our research endeavors to rectify these deficiencies through the implementation of Large Language Models (LLMs)\cite{yuTemporalDataMeets2023}.

For instance as shown in Figure \ref{fig:yelp}, consider the representation of information on a Yelp business POI depicted on a card. Previous endeavors predominantly relied on numeric data, such as latitude, longitude, and star ratings, alongside the utilization of attributes and categories as dummy variables. However, the incorporation of comments into mobility models proved challenging.

LLMs exhibit a proficiency in comprehending natural language, offering the potential to extract meaningful insights from POI text data. Leveraging this capability, the derived outputs can be seamlessly assimilated into mobility models, augmenting their intelligence and contextual relevance.

We conducted an in-depth analysis of select models that demonstrated strong performance in previous research efforts. Notably, we examined AGCRN, a traffic flow prediction model based on an adjacency matrix, and the Spatiotemporal Graph Convolutional Network (STGCN). Integrating STGCN model with large language models, we meticulously scrutinized the dataset we collected ourselves, yielding comprehensive and noteworthy results

\section{Related Works}
In this section, we discuss the relevant research that has been carried out in the relevant field.

Regarding the classical spatio‑temporal modeling methods, the spatio‑temporal prediction method based on the deep learning framework has been proposed. This method can be used to construct a continuous spatio‑temporal prediction framework based on discrete spatio‑temporal data points, which decomposes the spatio‑temporal process through the sum of the product of the basis functions, and then maps the spatio‑temporal data to the regular network, so as to establish complete spatio‑temporal data.

Another way to map discrete variable values to continuous variable values is to employ neural network embedding. Embedding refers to mapping discrete variables to continuous variables, and the embedding of neural networks can reduce the dimensionality of data, which is more conducive to the classification of datasets. The embedding of neural networks can generally be used as input data for deep learning frameworks, allowing for a clearer understanding of the correspondence between data and classification categories.

There is already a graph-based deep learning approach to demand forecasting for bike-sharing. The model used in this method is called B-MRGNN. The model consists of two main parts: a multi-relationship graph model and a convolutional neural network for multi-relationship graphs. It allows cross-modal information sharing and integrates dependencies from the spatiotemporal domain through a network connection between multiple modal plots, and performs well on multiple datasets.

At the same time, another model for the spatiotemporal long-term dependence of bike-sharing prediction tasks has been widely adopted: spatiotemporal convolutional network (STGCN). Instead of using convolutional layers and recurrent layers to build a more traditional neural network, it directly builds a complete convolutional neural network to reduce the number of parameters and improve the speed of model training.

\section{Approach}

\begin{figure}[t]
\begin{center}
\fbox{
   \includegraphics[width=0.65\linewidth]{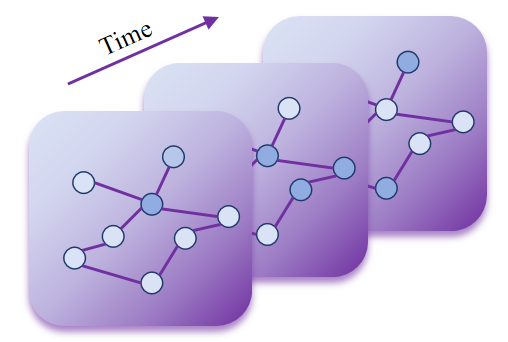}
}
\end{center}
   \caption{Graph-structured traffic data }
\label{fig:graph}
\end{figure}

\subsection{Bike Sharing Demand Prediction on Road Graphs}
Bike sharing demand forecasting is one kind of traffic forecast -- a typical time-series prediction problem. 

\textit{Problem (Bike Sharing Demand Prediction)}: This research aims to predict the most likely shared bike demand in the next $H$ time steps given the previous $M$ demand observations\cite{liangBikeSharingDemand2022a} as well as other languages data,
\begin{equation}
    \begin{split}
&\hat{d}_{t+1},...,\hat{d}_{t+H} = \\
    &\mathop{\arg\max}\limits_{d_{t+1},..,v_{t+H}} log\ P(d_{t+1},...,d_{t+H}|d_{t-M+1},...,d_t,\mathbf L)
    \end{split}
\end{equation}
where $d_t \in \mathbb R^n$ is an observation vector of $n$ shared bike stations at time step $t$, each element of which records historical observation for a single bike sharing node. And $\mathbf L$ is the embedding vector of the languages data for a single bike sharing node.

In the present study, we establish the traffic network as a graph, with a specific emphasis on structured traffic time series analysis. The observation $d_t$ is inherently interconnected rather than independent, forming pairwise connections within the graph structure. Consequently, the data point $d_t$ can be conceptualized as a graph signal defined on an undirected graph denoted as $\mathbf G$, featuring weights $w_{ij}$ as delineated in Figure \ref{fig:graph}.

At the $t$-th time step, within the graph $\mathbf G_t = (\mathbf V_t, \mathbf L, \bf\epsilon, W)$, $\mathbf{V_t}$ represents a finite set of vertices, corresponding to observations from $n$ bike-sharing stations within the traffic network. Additionally, $\mathbf{L}$ is a vector encapsulating language data associated with Points of Interest (POIs) around bike-sharing stations. The parameter $\epsilon$ denotes a set of edges, elucidating the interconnectedness between stations, while $\mathbf{M} \in \mathbb R^{n\times n}$ signifies the weighted adjacency matrix of $\mathbf{G}_t$ \cite{amatoNovelFrameworkSpatiotemporal2020}.

\subsection{Embeddings}
\begin{figure}[t]
\begin{center}
\fbox{
   \includegraphics[width=0.95\linewidth]{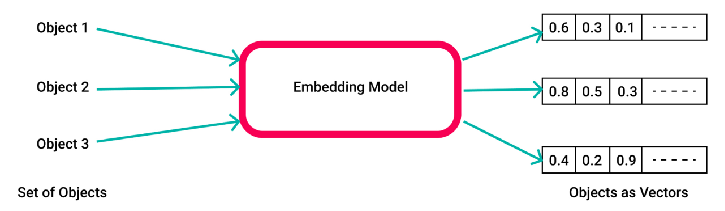}
}
\end{center}
   \caption{Example of Embeddings. Embeddings are numerical representations of concepts converted to number sequences, which make it easy for computers to understand the relationships between those concepts. }
\label{fig:embedding}
\end{figure}

An embedding refers to the mapping of a discrete, categorical variable onto a vector of continuous numbers, as illustrated in Figure \ref{fig:embedding}. Within the realm of neural networks, embeddings represent low-dimensional, learned continuous vector representations of discrete variables. The utility of neural network embeddings stems from their ability to effectively reduce the dimensionality of categorical variables, providing meaningful representations for categories within the transformed space \cite{koehrsenNeuralNetworkEmbeddings2018}.

These embeddings prove valuable in the context of natural language and code processing, given their ease of consumption and comparability by other machine learning models and algorithms, such as clustering or search functionalities. In this study, we employ the OpenAI GPT-4 Embeddings API to convert text into a $1536$-dimensional vector, utilized as the language data $\mathbf L$.

\begin{figure*}[t]
\begin{center}
\fbox{
\includegraphics[width=0.8\linewidth]{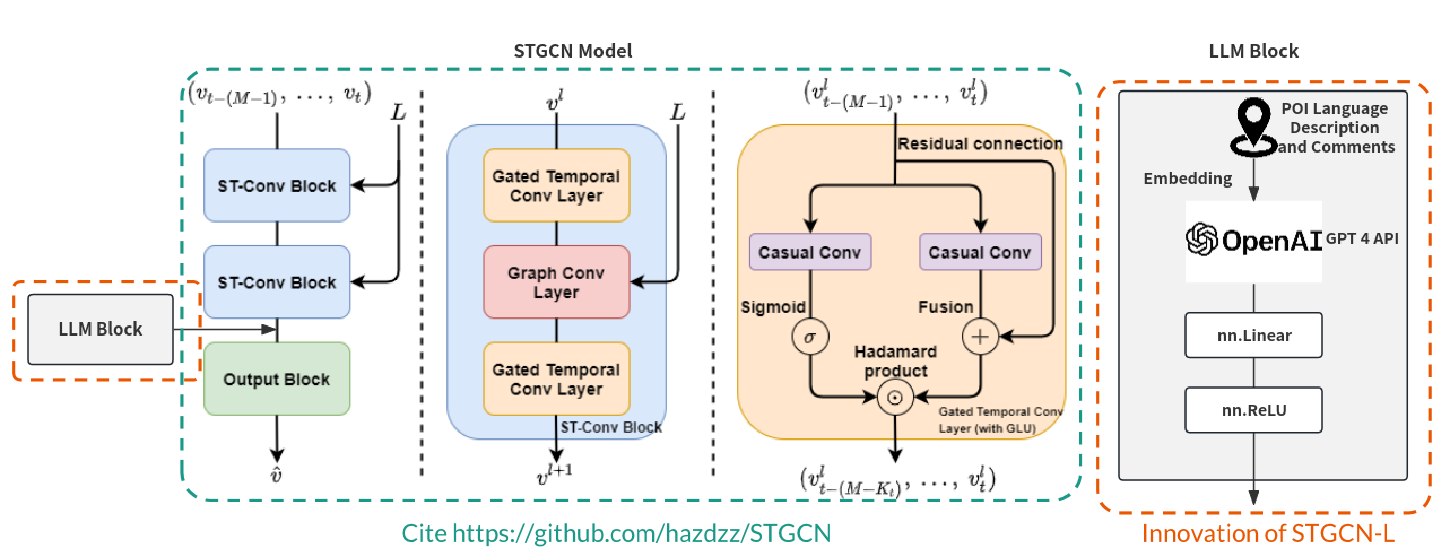}
}
\end{center}
   \caption{Architecture of spatio-temporal graph convolutional networks combined with Large Language Model. The framework STGCN-L consists of two spatio-temporal convolutional blocks (ST-Conv blocks), a LLM block as an encoder and a fully-connected output layer in the end. Each ST-Conv block contains two temporal gated convolution layers and one spatial graph convolution layer in the middle. The residual connection and bottleneck strategy are applied inside block.}
\label{fig:structure}
\end{figure*}

\subsection{STGCN}
Spatio-temporal graph convolutional networks, is a deep learning model to do traffic forecasting works. As shown in the green box of Figure \ref{fig:structure}, STGCN is composed of several spatio-temporal convolutional blocks, each of which is formed as a "sandwich" structure with two gated sequential convolution layers and one spatial graph convolutional layer in between.\cite{yuSpatioTemporalGraphConvolutional2018}

\subsection{STGCN-L}
In this section, we augment the proposed Spatio-Temporal Graph Convolutional Network (STGCN) architecture by introducing a Large Language Model (LLM) block. As depicted in Figure \ref{fig:structure}, the LLM block is strategically positioned between the second ST-Conv Block and the output layer. Within the LLM block, the language data associated with Points of Interest (POIs) undergoes conversion into 1536-dimensional vectors using the OpenAI GPT-4 embeddings API, thereby serving as spatial features for each node.

To evaluate the performance of our model, we employ the $L2$ loss metric. Consequently, the loss function of STGCN-L for bike-sharing demand is formulated as follows:
\begin{equation}
    L(\hat v;L_{\theta},L)=\sum_t||\hat v(v_{t-M+1},...,v_t,L_{\theta},L)-v_{t+1}||^2
\end{equation}
where $L_{\theta}$ are all trainable parameters in the model; $L$ are embeddings in the model; $v_{t+1}$ is the ground truth and $\hat v(\cdot )$ denotes the model's predicition.

\section{Experiments}

\subsection{Data sets}

\textbf{a. Map Boundaries}

We get the map boundaries from OPEN Philadelphia 2010 Census Block Groups which divides the city into 1336 regions.

\textbf{b. Bike Sharing Demand Data}

\begin{figure}[h]
	\centering
	\begin{minipage}{0.49\linewidth}
		\centering
		\includegraphics[width=0.9\linewidth]{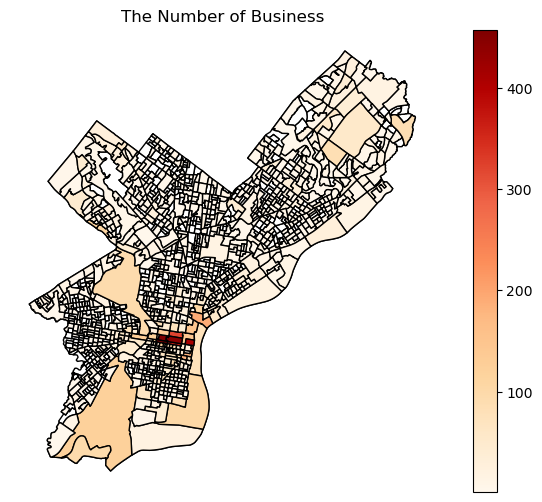}
		\caption{Business POI in City Philadelphia}
		\label{fig:business}
	\end{minipage}
	\begin{minipage}{0.49\linewidth}
		\centering
		\includegraphics[width=0.9\linewidth]{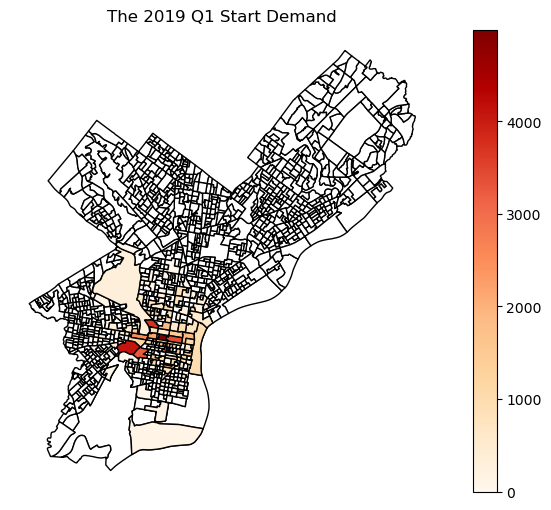}
		\caption{Bike Sharing Demand of 2019Q1 in City Philadelphia}
		\label{fig:demand}
	\end{minipage}
\end{figure}

We get the Bike Sharing Dataset of Philadelphia From 2019Q1 to 2023Q1. The Figure \ref{fig:demand} shows the distribution of Bike sharing demand of 2019Q1 in City Philadelphia.

\textbf{c. Yelp Open Data}

The Yelp dataset constitutes a subset encompassing Yelp businesses, reviews, and user data, designed for utilization in personal, educational, and academic contexts. Presented in JSON file format, it serves as a valuable resource for instructing students on database principles, facilitating the comprehension of Natural Language Processing (NLP), and offering sample production data for individuals acquiring skills in mobile application development. In this study, we specifically extract the Yelp Points of Interest (POI) data pertaining to the city of Philadelphia, shaping it into a language dataset. The distribution of business POIs in Philadelphia, illustrated in Figure \ref{fig:business}, is noted for its pronounced imbalance.

\subsection{Data Processing}
We obtain the reviews and descriptions of each business, treating them as embeddings that serve as features for each business Point of Interest (POI). Subsequently, we compute the average vector of all embeddings within a given region, considering this as the feature representation for that region.

\begin{figure}[t]
\begin{center}
\fbox{
   \includegraphics[width=0.65\linewidth]{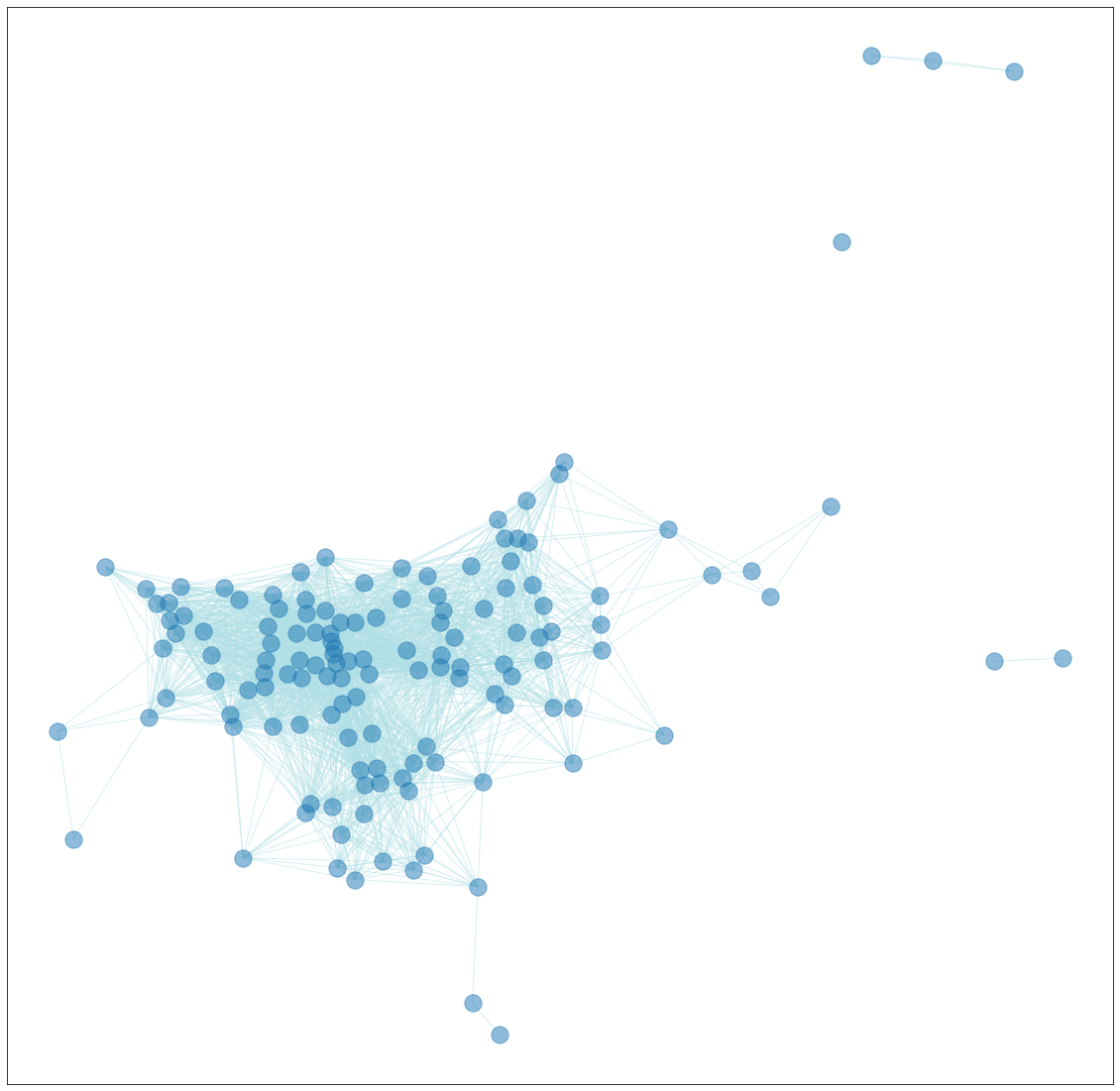}
}
\end{center}
   \caption{The Bike Sharing Traffic Network Graph}
\label{fig:adjacency}
\end{figure}

It is imperative to note that not all regions possess both Bike Sharing Demand data and Yelp Language data. Consequently, we identify and extract 136 regions that do not exhibit any missing data. Defining the traffic network graph of Philadelphia with 136 nodes and 1,536 features for each node, illustrated in Figure \ref{fig:adjacency}, we compute the adjacency matrix for the graph, utilizing the true distance as edge weight. Nodes beyond a distance of 160 km are designated as unconnected. Furthermore, we temporally disaggregate the bike sharing demand data into hourly intervals, resulting in a time series dataset comprising 43,832 steps.

\subsection{Experimental Settings}
All experiments are compiled and tested on a Window PC (CPU: 12th Gen Intel(R) Core(TM) i5-12400 @ 2.5GHz, GPU: NVIDIA GeForce RTX 3060 Ti).We mainly use pytorch as the framework for deep learning.
\subsection{Results}
\subsubsection{STGCN}
The GPU usage for the STGCN model is set to True, the input time step is set to 12, the output time step is set to 3, the epochs are set to 1000, the batch\_size = 50.Its average MSE and MAE are 0.516 and 0.615 respectively. 
The loss curves and MAE curves of the training and test sets are shown below, and it can be seen that the loss curve of the training set decreases slowly after 40 epochs, while the loss curve of the test set basically reaches a relatively stable result within 10 epochs, and thereafter the amplitude of the bands does not exceed 0.04 within 90 epochs.

\subsubsection{STGCN-L}
Compared to the STGCN model, the STGCN-L model utilizes the LLM to add new features to each node in hopes of improving training. Other parameter settings are basically identical to the STGCN method. Its average MSE and MAE are 0.506 and 0.595 respectively. the loss curves and MAE curves of the training and test sets are shown in the following figure, which shows that the overall trend of the loss is consistent with that of the STGCN, but MSE and MAE are smaller when stabilized.

\subsubsection{AGCRN}
The batch\_size is set to 64, the lag is set to 12, the ratio of training set, test set and validation set is set to 6:2:2, and the initial learning rate is set to 0.003. For other more parameter settings, please see the Feicheng\_AGCRN.conf file.
The MAE and MSE of the AGCRN method are 0.64 and 0.30 respectively.The loss curves for the training and validation sets are shown below.
We find that the method converges faster and the loss curve on the validation set is close to smooth within 5 epochs.

\begin{figure}[t]
        \centering
        \subfloat[AGCRN]{\includegraphics[width=0.3\linewidth]{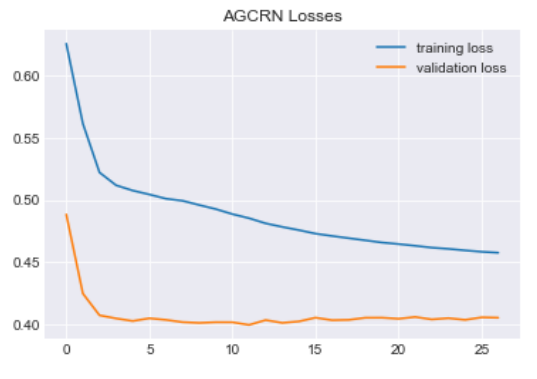}}
	\subfloat[STGCN]{\includegraphics[width=0.3\linewidth]{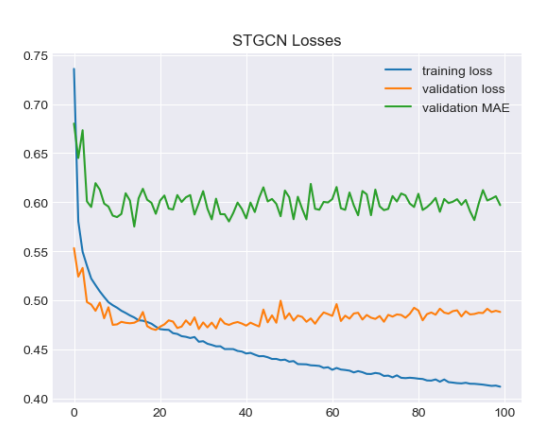}}
        \subfloat[STGCN-L]{\includegraphics[width=0.3\linewidth]{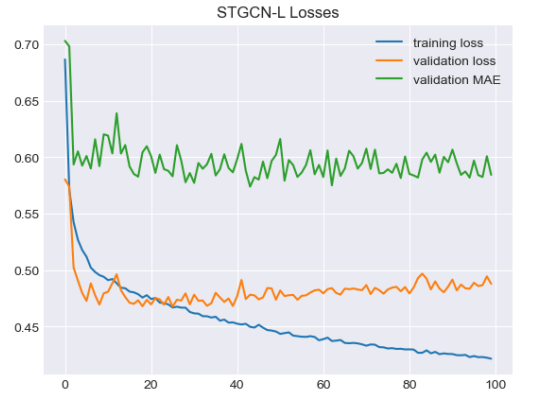}}
	\caption{Experiment Training Losses}
	\label{fig:results}
\end{figure}

\begin{table}
\begin{center}
\scalebox{0.9}{
    \begin{tabularx}{0.8\linewidth} { 
      | >{\raggedright\arraybackslash}X 
      | >{\centering\arraybackslash}X 
      | >{\raggedleft\arraybackslash}X | }
     \hline
     \textbf{Models} & \textbf{MSE} & \textbf{MAE} \\
     \hline
     STGCN\-L  & \textbf{0.506}  & 0.595  \\
    \hline
     STGCN  & 0.516  & 0.615  \\
    \hline
     AGCRN  & 0.64  & \textbf{0.30}  \\
    \hline
    \end{tabularx}
    }
\end{center}
\caption{Results}
\end{table}

\section{Conclusion}
In comparing the outcomes of the aforementioned three methodologies, it is observed that the STGCN-L model exhibits the lowest Mean Squared Error (MSE), whereas the AGCRN model demonstrates the smallest Mean Absolute Error (MAE), with the STGCN's performance falling in between. Theoretically, AGCRN introduces adaptive graph convolution and a recurrent network, enhancing its capacity to capture intricate relationships in spatio-temporal data and improve the modeling of nonlinear and dynamic changes. Consequently, one would anticipate AGCRN to outperform STGCN. However, empirically, the MSE of AGCRN is the highest among the three methods. This discrepancy may be attributed to the presence of prominent outliers in the data, necessitating a more nuanced feature representation to achieve precise descriptions and predictions.

Additionally, AGCRN is presumed to entail a greater number of hyperparameters to be fine-tuned compared to STGCN, potentially rendering it more susceptible to tuning-induced variations in performance. Subsequent to the incorporation of the Large Language Model (LLM) Block, the performance optimization of STGCN exhibits a more modest improvement. This suggests that the inclusion of mass evaluation as a new feature contributes somewhat to the understanding of inter-site traffic dynamics. However, it is acknowledged that other pivotal factors, such as weather conditions, may exert more substantial influences. In forthcoming endeavors, efforts will be directed towards augmenting prediction accuracy by introducing additional features and exploring alternative models.
\section{Acknowledgement}
We would like to express our sincere gratitude to the MIT Transit Lab, under the guidance of Dr. Jinhua Zhao, Dr. Shenhao Wang, and the invaluable contributions of PhD student Dingyi Zhuang and Qingyi Wang. Their expertise, mentorship, and collaborative efforts have been instrumental in the success of this research. Their insightful guidance and support have significantly enriched the depth and quality of our work. I am truly appreciative of their dedication and commendable contributions to the advancement of our research endeavors.

{\small
\bibliographystyle{ieee_fullname}
\bibliography{main}

\begin{thebibliography}{1}\itemsep=-1pt

\bibitem{amatoNovelFrameworkSpatiotemporal2020}
Federico Amato, Fabian Guignard, Sylvain Robert, and Mikhail Kanevski.
\newblock A novel framework for spatio-temporal prediction of environmental data using deep learning.
\newblock {\em Scientific Reports}, 10(1):22243, Dec. 2020.

\bibitem{kipf2016semi}
Thomas~N Kipf and Max Welling.
\newblock Semi-supervised classification with graph convolutional networks.
\newblock {\em arXiv preprint arXiv:1609.02907}, 2016.

\bibitem{koehrsenNeuralNetworkEmbeddings2018}
Will Koehrsen.
\newblock Neural {{Network Embeddings Explained}}.
\newblock https://towardsdatascience.com/neural-network-embeddings-explained-4d028e6f0526, Oct. 2018.

\bibitem{liangBikeSharingDemand2022a}
Yuebing Liang, Guan Huang, and Zhan Zhao.
\newblock Bike {{Sharing Demand Prediction}} based on {{Knowledge Sharing}} across {{Modes}}: {{A Graph-based Deep Learning Approach}}, Mar. 2022.

\bibitem{liangJointDemandPrediction2022}
Yuebing Liang, Guan Huang, and Zhan Zhao.
\newblock Joint {{Demand Prediction}} for {{Multimodal Systems}}: {{A Multi-task Multi-relational Spatiotemporal Graph Neural Network Approach}}.
\newblock {\em Transportation Research Part C: Emerging Technologies}, 140:103731, July 2022.

\bibitem{yuSpatioTemporalGraphConvolutional2018}
Bing Yu, Haoteng Yin, and Zhanxing Zhu.
\newblock Spatio-{{Temporal Graph Convolutional Networks}}: {{A Deep Learning Framework}} for {{Traffic Forecasting}}.
\newblock In {\em Proceedings of the {{Twenty-Seventh International Joint Conference}} on {{Artificial Intelligence}}}, pages 3634--3640, July 2018.

\bibitem{yuTemporalDataMeets2023}
Xinli Yu, Zheng Chen, Yuan Ling, Shujing Dong, Zongyi Liu, and Yanbin Lu.
\newblock Temporal {{Data Meets LLM}} -- {{Explainable Financial Time Series Forecasting}}, June 2023.

\end{thebibliography}
}
\newpage
\appendix
\section{GCN}
Currently, most graph neural network models have a somewhat universal architecture in common. We refer to these models as Graph Convolutional Networks (GCNs); convolutional, because filter parameters are typically shared over all locations in the graph (or a subset thereof as in Duvenaud et al., NIPS 2015).

For these models, the goal is then to learn a function of signals/features on a graph $\mathcal G=(\mathcal V,\mathcal E)$
 which takes as input:
\begin{itemize}
    \item A feature description $x_i$ for every node $i$; summarized in a $N\times D$ feature matrix $X $($N$: number of nodes, $D$: number of input features)
    \item A representative description of the graph structure in matrix form; typically in the form of an adjacency matrix $A$ (or some function thereof)
\end{itemize}

and produces a node-level output $Z$ (an $N\times F$ feature matrix, where $F$ is the number of output features per node). Graph-level outputs can be modeled by introducing some form of pooling operation (see, e.g. Duvenaud et al., NIPS 2015).

Every neural network layer can then be written as a non-linear function
$$
H^{(l+1)}=f(H^{(l)},A)
$$
with $H(0)=X$ and $H(L)=Z$ (or $z$ for graph-level outputs), $L$ being the number of layers. The specific models then differ only in how $f(\cdot,\cdot)$ is chosen and parameterized.

As an example, let's consider the following very simple form of a layer-wise propagation rule:
$$f(H^{(l)},A)=\sigma(AH^{(l)}W^{(l)})$$
where $W^{(l)}$ is a weight matrix for the $l$-th neural network layer and $\sigma(\cdot)$ is a non-linear activation function like the \textbf{ReLU}. Despite its simplicity this model is already quite powerful (we'll come to that in a moment).

But first, let us address two limitations of this simple model: multiplication with $A$
means that, for every node, we sum up all the feature vectors of all neighboring nodes but not the node itself (unless there are self-loops in the graph). We can "fix" this by enforcing self-loops in the graph: we simply add the identity matrix to $A$.

The second major limitation is that $A$ is typically not normalized and therefore the multiplication with $A$ will completely change the scale of the feature vectors (we can understand that by looking at the eigenvalues of $A$). Normalizing $A$ such that all rows sum to one, i.e. $D^{-1}A$, where $D$ is the diagonal node degree matrix, gets rid of this problem. Multiplying with $ D^{-1}A$ now corresponds to taking the average of neighboring node features. In practice, dynamics get more interesting when we use a symmetric normalization, i.e. $ D^{-1/2}A^{-1/2}$ (as this no longer amounts to mere averaging of neighboring nodes). Combining these two tricks, we essentially arrive at the propagation rule introduced in Kipf \& Welling (ICLR 2017):
 $$f(H^{(l)},A)=\sigma(AH^{(l)}W^{(l)})=\sigma(\hat D^{-1/2}\hat A\hat D^{-1/2}H^{(l)}W^{(l)})$$
with $\hat A=A+I$, where $I$ is the identity matrix and $\hat D$ is the diagonal node degree matrix of $\hat A$.\cite{kipf2016semi}

\end{document}